\documentclass{jpsj2}
\usepackage{rotating}

\def\runauthor{Online-Journal Subcommittee}

\title{%
Decay Properties of $^{266}$Bh and $^{262}$Db Produced in the $^{248}$Cm + $^{23}$Na Reaction
}

\author{%
Kosuke \textsc{Morita}$^{1}$\thanks{E-mail address: morita@ribf.riken.jp}, 
Kouji \textsc{Morimoto}$^{1}$,
Daiya \textsc{Kaji}$^{1}$,
Hiromitsu \textsc{Haba}$^{1}$, 
Kazutaka \textsc{Ozeki}$^{1}$,
Yuki \textsc{Kudou}$^{1}$,
Nozomi \textsc{Sato}$^{1,2}$\thanks{Present address: Japan Atomic Energy Agency},
Takayuki \textsc{Sumita}$^{3}$,
Akira \textsc{Yoneda}$^{1}$,
Takatoshi \textsc{Ichikawa}$^{1}$\thanks{Present address: Yukawa Institute for Theoretical Physics, Kyoto University}, 
Yasuyuki \textsc{Fujimori}$^{4}$, 
Sin-ichi \textsc{Goto}$^{5}$,
Eiji \textsc{Ideguchi}$^{6}$,
Yoshitaka \textsc{Kasamatsu}$^{7}$\thanks{Present address: Nishina Center for Accerelator-Based Science, RIKEN},
Kenji \textsc{Katori}$^{1}$,
Yukiko \textsc{Komori}$^{8}$,
Hiroyuki \textsc{Koura}$^{7}$,
Hisaaki \textsc{Kudo}$^{9}$,
Kazuhiro \textsc{Ooe}$^{8}$,
Akira \textsc{Ozawa}$^{10}$,
Fuyuki \textsc{Tokanai}$^{4}$,
Kazuaki \textsc{Tsukada}$^{7}$,
Takayuki \textsc{Yamaguchi}$^{11}$,
 and 
Atsushi \textsc{Yoshida}$^{1}$
}

\inst{%
$^{1}$Nishina Center for Accelerator-Based Science, RIKEN, Wako-shi, Saitama 351-0198 \\
$^{2}$Department of Physics, Tohoku University, Aoba-ku, Sendai 980-8578 \\
$^{3}$Faculty of Science and Technology, Tokyo University of Science, Noda, Chiba 278-8510 \\
$^{4}$Department of Physics, Yamagata University, Yamagata-shi, Yamagata 990-8560 \\
$^{5}$Center for Instrumental Analysis, Niigata University, Ikarashi, Nishi-ku, Niigata 950-2181 \\
$^{6}$Center for Nuclear Study, University of Tokyo, Wako-shi, Saitama 351-0198 \\
$^{7}$Japan Atomic Energy Agency, Tokai, Ibaraki 319-1195 \\
$^{8}$ Department of Chemistry, Osaka University, Toyonaka-shi, Osaka 560-0043 \\
$^{9}$Department of Chemistry, Niigata University, Ikarashi, Nishi-ku, Niigata 950-2181 \\
$^{10}$University of Tsukuba, Tsukuba, Ibaraki 305-8571 \\
$^{11}$Department of Physics, Saitama University, Saitama-shi, Saitama 338-8570 \\
}

\recdate{\today}

\abst{%

Decay properties of an isotope $^{266}$Bh and its daughter nucleus $^{262}$Db produced by the $^{248}$Cm($^{23}$Na, 5\textit{n}) reaction were studied by using a gas-filled recoil separator coupled with a position-sensitive semiconductor detector.  $^{266}$Bh was clearly identified from the correlation of the known nuclide, $^{262}$Db. The obtained decay properties of $^{266}$Bh and $^{262}$Db are consistent with those observed in the $^{278}$113 chain, which provided further confirmation of the discovery of $^{278}$113 .
}

\kword{%
$^{266}$Bh, $^{262}$Db, $\alpha$-decay, spontaneous fission, gas-filled recoil ion separator, position-sensitive focal plane detector
}

\begin{document}

\maketitle

\section{Introduction}

The identification of the heaviest nuclides is very difficult because of their extremely small production cross sections. Very heavy nuclides usually form an $\alpha$-decay chain, and they can be conclusively assigned based on a genetic link to known nuclide(s). A nuclide, $^{266}$Bh, is the great-granddaughter of $^{278}$113 that is produced in the $^{209}$Bi + $^{70}$Zn reaction \cite{morita113-1, morita113-2}. Thus far, only five atoms have been assigned to $^{266}$Bh by direct production \cite{wilk00,Qin06}. Most of the heaviest odd-odd nuclei exhibit rather complicated decay properties. Their main decay modes are $\alpha$-decay, electron-capture (EC) and spontaneous fission (SF). In many cases, these modes coexist in one nucleus. Furthermore, $\alpha$-decay energies widely distribute because many low-lying states exist in their daughter nuclei. Therefore, it is difficult to obtain a clear assignment of heavy odd-odd nuclides such as $^{266}$Bh. Wilk \textit{et al.} \cite{wilk00} reported the production of $^{266}$Bh in the $^{249}$Bk + $^{22}$Ne reaction by the use of a helium gas-jet technique. They observed one decay chain assigned to one originating from $^{266}$Bh ($^{266}$Bh $\longrightarrow $ $^{262}$Db $\longrightarrow $ $^{258}$Lr $\longrightarrow $). Qin \textit{et al.} \cite{Qin06} studied the decay properties of $^{266}$Bh produced in the $^{243}$Am + $^{26}$Mg reaction by the use of the gas-jet technique. They observed four $\alpha$-decay chains originating from $^{266}$Bh. In the present work, we performed an experiment with the aim of obtaining an unambiguous assignment of $^{266}$Bh by a genetic link. For this purpose, we used a gas-filled recoil ion separator (GARIS) coupled to a position-sensitive focal plane detector (PSD). Using a rotating target of $^{248}$Cm, $^{266}$Bh was produced for the first time in the $^{248}$Cm + $^{23}$Na reaction. The main purpose of this work is to provide further confirmation of the production and identification of the isotope $^{278}$113.

\section{Experimental Procedure}

\begin{figure}[t]
\begin{center}
\includegraphics[width=14cm]{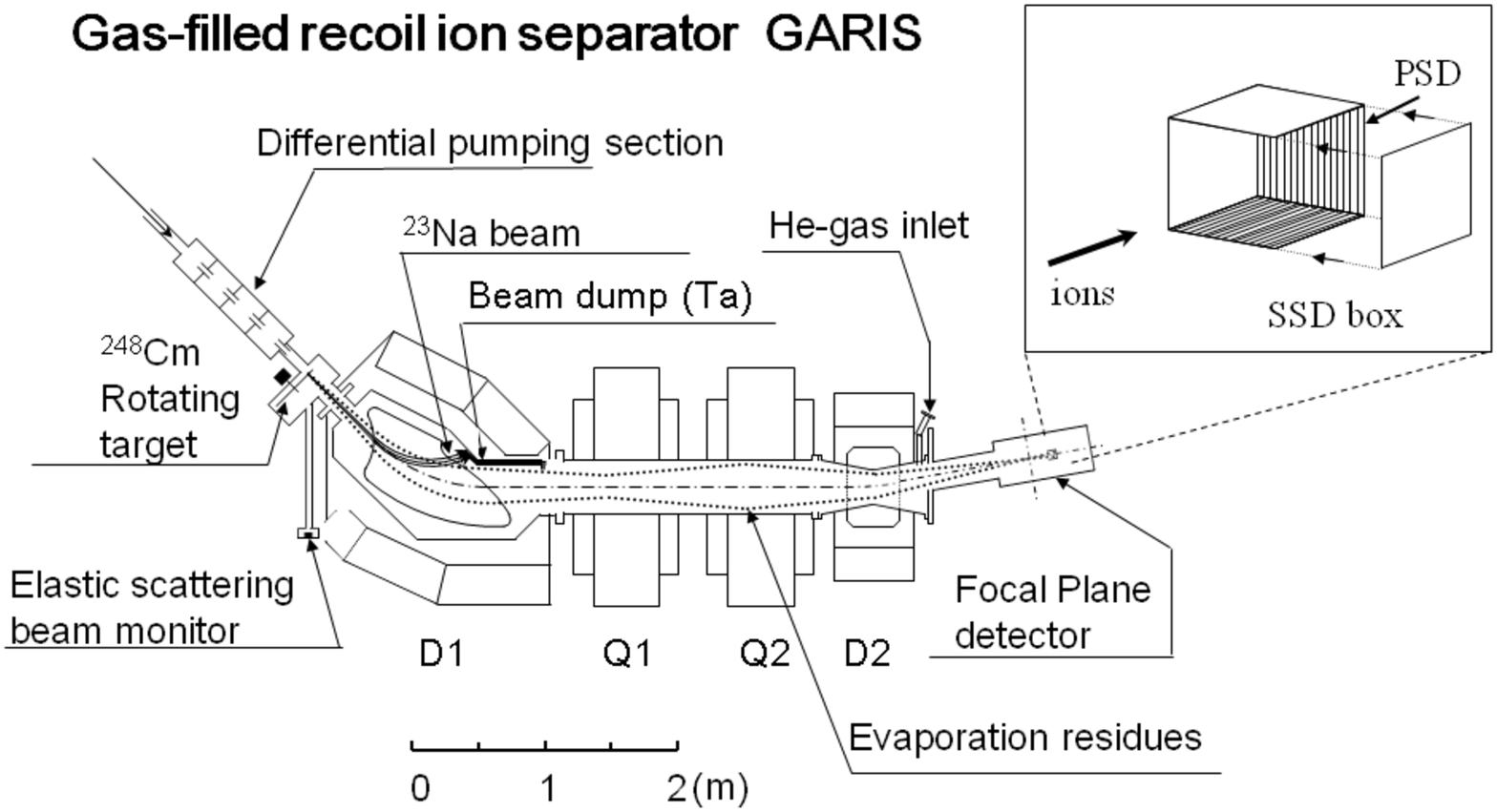}
\end{center}
\vspace{-1 cm}
\caption{Schematic view of the experimental setup.}
\label{f1}
\end{figure}

The experiment was performed at the RIKEN Linear Accelerator (RILAC) Facility.  A schematic view of the experimental setup is shown in Fig.~\ref{f1}.  

A $^{248}$Cm$_{2}$O$_{3}$ target having a thickness of 350 $\mu$g/cm$^{2}$ was prepared by electrodeposition onto a titanium backing foil having a thickness of 0.91 mg/cm$^{2}$.  Six pieces of the target were mounted on a rotating wheel having a diameter of 10 cm.  During irradiation, the wheel was rotated at 1000 rpm.  Although a maximum of eight pieces of the target could be mounted on the wheel, we used only six pieces because of a lack of the curium material.

A $^{23}$Na beam was extracted from RILAC.  Beam energies of 126, 130 and 132 MeV were used.  The corresponding beam energies at the middle of the target were 121, 124 and 126 MeV respectively.  The $^{23}$Na beam was pulsed with micro- and macro-pulse structures.  The micro-pulse structure was used to prevent the irradiation of the target frame and the two vacant target positions of the wheel. The micro-pulse structure was synchronized to the rotation of the wheel.  In the macro-pulse structure, the beam was ON for 3 s and then OFF for 3 s independently of the micro-pulse structure.  A logical AND signal of the micro- and macro-pulse structures was sent to the beam pulsing system of the accelerator.  All the measurements were performed  only in the beam OFF period.  The counting duty in the macro beam OFF period was 100\%, while that in the macro beam ON period was 45\%.  The beam duty factor was 27.5\%.  The typical beam intensity was 1 particle-$\mu$A on average, and it was 4.4 particle-$\mu$A in the micro beam ON period.

GARIS was used to collect evaporation residues (ERs) and separate them from the beam particles and other unwanted charged particles.  GARIS was filled with helium gas at a pressure of 33 Pa.  Prior to the present experiment, we measured the optimum magnetic rigidity (B$\rho $) for $^{265}$Sg produced in the $^{248}$Cm($^{22}$Ne, 5\textit{n})$^{265}$Sg reaction, which differs by only one proton from  $^{266}$Bh. The highest yield of $^{265}$Sg was obtained at 2.07 Tm. Thus, B$\rho $ of GARIS was set at 2.07 Tm for beam energies of 132 and 130 MeV.  At beam energies of 126 MeV and a part of 132 MeV, B$\rho $ was set at 2.19 Tm to reduce the counting rate of the focal plane detector to one half.

ERs were transported to the focal plane of GARIS where the PSD was set.  The arrangement of the focal plane detector is shown in Fig.~\ref{f1}.  The effective area of the PSD is 60 mm $\times$ 60 mm.  It comprises sixteen strip detectors having a width of 3.75 mm each, and each strip detector has position sensitivity.  At a backward of the PSD, semiconductor detectors (SSDs) having the same effective area as the PSD were set in a box shape to detect particles emitted in backward angle from the ERs and their decay daughters.

Because the mass of the projectile is much smaller than that of target for the reaction studied in the present work (asymmetric reaction), the kinetic energy of the ERs is as low as 10 MeV at the focal plane. Consequently, the depth of the implantation into the detector is approximately 1.5 $\mu$m.  Therefore, if an $\alpha$-particle is emitted at a backward angle, the energy  deposition to the PSD is very small, i.e., a few hundred keV depending on the angle of emission.  The position resolving power of the PSD is proportional to the energy deposited in it when the total energy of the $\alpha$-particle is measured by the sum of PSD and SSD (PSD + SSD event); therefore, in this case, the position resolution becomes poor.  A position window should be selected carefully when searching for a correlation event.

\begin{figure}[t]
\begin{center}
\resizebox{0.8\textwidth}{!}{\includegraphics*{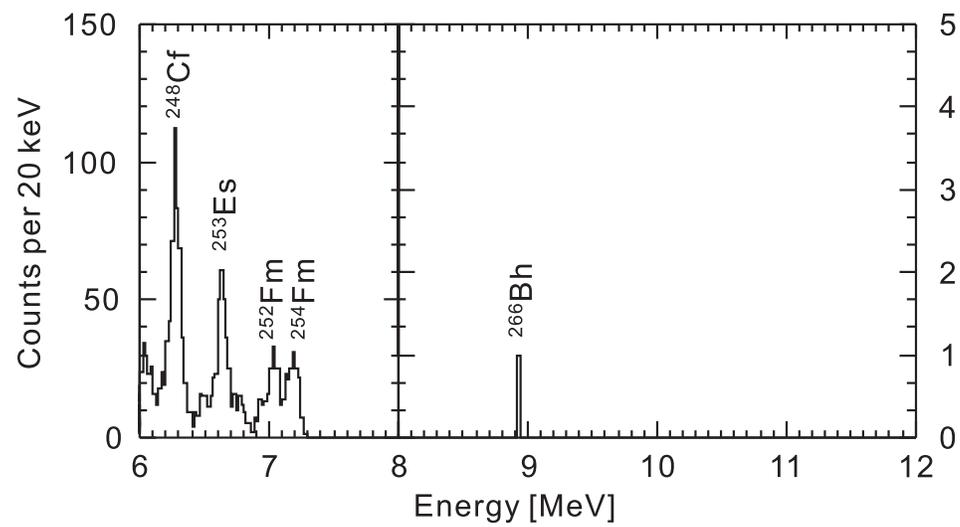}}%
\end{center}
\caption{Example of a singles spectrum for run161. The 7th strip among 16 (0-15) in the PSD was used.  The measurement time was 16.4 h for a beam dose of 3.1 $\times$ 10$^{17}$.}
\label{f2}
\end{figure}

The total counting rate of the focal plane detector was approximately 3 $\times$ 10$^{4}$/s in the beam ON period, while that was 5--10/s in the beam OFF period.  Because of the high implantation rate to the PSD in the beam ON period, the PSD was gradually damaged during the measurement.  Therefore, the energy resolution of the PSD gradually worsened with time.  We replaced the PSD three times during the long experimental period of one and a half months to remove this problem.  Energy calibration of the focal plane detectors was performed using long-lived transfer products implanted in the PSD simultaneously with the ERs of interest.  The alpha lines used in the calibration were 6.258 MeV ($^{248}$Cf), 7.039 MeV ($^{252}$Fm) and 7.192 MeV ($^{254}$Fm).  The energy resolution was measured simultaneously. Occasionally, we also used an $^{241}$Am source  to check the energy calibration and resolution.  An example of a singles spectrum is shown in Fig.~\ref{f2}.  The spectrum was obtained from one strip in one run (run161) with a beam dose of 3.1 $\times$ 10$^{17}$.  The measurement time was 16.4 h.  In the energy region below 8 MeV, $\alpha$ lines from long-lived transfer products are observed, while in the region from 8 to 12 MeV, only one event that was  assigned to an $\alpha$ decay of $^{266}$Bh was detected.

We performed offline analyses to search for mother-daughter correlations, i.e. $\alpha$-$\alpha$ and $\alpha$-SF events, based on the position in PSD, energy, and time difference.  Because we only considered data from the beam OFF period, a signal indicating the implantation of ER was not recorded.  Therefore, the time difference between the ER implantation and its decay event was not obtained.

The average counting rate of particles having energies greater than 50 MeV for each strip of the PSD was 2.7 $\times$ 10$^{-4}$/s.  Considering a positional window of 2 mm, which is a typical position resolution for a true coincidence event for a full energy deposited $\alpha$-particle and SF, $\pm$ 1 mm, the counting rate in the positional window is calculated to be 9.3 $\times$ 10$^{-6}$/s.  Considering a time window of 300 s, the probability of an accidental correlation is calculated to be 2.8 $\times$ 10$^{-3}$ per event.  This low value of accidental probability was realized by using a combination of GARIS and the PSD.  The reduction in the number of transfer products and target recoils at the focal plane that is achieved using the physical pre-separator, GARIS and the high spatial resolving power of the PSD enables a low background $\alpha$-SF correlation measurements. The $\alpha$-SF correlation measurement would be difficult to perform using only the gas-jet technique. 

\section{Results and Discussion}

We have assigned total of 32 correlations to true ones.  All the correlated events are listed in Table I.  The event ID\#, beam energies from the accelerator, strip number of the PSD, decay energies and the energy resolutions (in full width at half maximum, FWHM), position difference (in mm), grouping in correlation map mentioned below and assignments for each event are listed. Four of the events are $\alpha_{1}$-$\alpha_{2}$-$\alpha_{3}$ correlations (ID\#1--ID\#4), one is an $\alpha_{1}$-$\alpha_{2}$-SF (ID\#5), nine are $\alpha_{1}$-$\alpha_{2}$ (ID\#6--ID\#14) and eighteen are $\alpha_{1}$-SF correlations (ID\#15--ID\#32).  The maximum correlation time was set to 300 s.  The position window was set to $\pm$ 2 mm for the correlation events measured only in the PSD (PSD events).  The position window was properly selected for every PSD + SSD event.  The superscript 's' in the energy indicated that an event was a PSD + SSD event.
A two-dimensional representation of the time- and position-correlated events is shown in Fig.~\ref{f3}.  The horizontal and vertical axes represent the mother and daughter energy, respectively.  The lower and upper panels (Figs.~\ref{f3}-a and ~\ref{f3}-b) show the $\alpha$-$\alpha$ and $\alpha$-SF correlations, respectively.

The excitation energies of the compound nucleus, $^{271}$Bh, at the middle of the target were calculated to be 44.4, 48.1 and 49.9 MeV for beam energies of 126, 130 and 132 MeV, respectively, by using a theoretical mass for $^{271}$Bh \cite{moller95}.  The energies were selected to maximize the 5\textit{n} evaporation channel in the $^{248}$Cm + $^{23}$Na reaction in order to produce the isotope $^{266}$Bh.  Because the ER cross sections of these asymmetric reactions in the heaviest region exhibit a broad peaks around the optimum energies that produce the maximum yields, the isotope $^{267}$Bh, which is a product of the 4\textit{n}  evaporation channel of the reaction, could be produced in considerable ammounts, especially at lower excitation energies.  We focused on these two isotopes in the analyses.  The direct productions of $^{262}$Db and $^{263}$Db, which are the $\alpha$-decay daughters of $^{266}$Bh and $^{267}$Bh by the $\alpha$ + 5\textit{n} and $\alpha$ + 4\textit{n} evaporation channels, respectively, were not considered because of the reaction in the sub-barrier energy region.

\subsection{$\alpha$-$\alpha$ correlations}

In the correlations shown in Fig.~\ref{f3}-a, eight points are plotted in Group C (ID\#1--ID\#4, ID\#10--ID\#13). The mother energies of these events range from 8.40 to 8.74 MeV, and those of the corresponding daughters range from 8.57 to 8.80 MeV. We assigned these events to the $^{262}$Db $\longrightarrow $ $^{258}$Lr $\longrightarrow$ decay.  The average time differences between mother and daughter decays, i.e. the mean life of the daughter nuclide, was calculated to be 5.8 $^{+ 3.2}_{- 1.5}$ s.  The half-life T$_{1/2}$ is deduced to be 4.0 $^{+ 2.2}_{- 1.0}$ s.  The obtained T$_{1/2}$ in this work agrees very well with the adopted value of  $^{258}$Lr, 3.92 $^{+ 0.35}_{- 0.42}$ s \cite{dressler99}. The adopted $\alpha$-decay energies of $^{262}$Db are 8.45 $\pm$ 0.02, 8.53 $\pm$ 0.02 and 8.67 $\pm$ 0.02 MeV, and those of $^{258}$Lr are 8.565 $\pm$ 0.025, 8.595 $\pm$ 0.010, 8.621 $\pm$ 0.010, and 8.654 $\pm$ 0.010 MeV \cite{dressler99}.  It should be noted that the $\alpha$ energies obtained in this work are approximately 40--60 keV higher than the adopted energies. This is most probably attributable to the summing of $\alpha$-energy and a conversion electron or $\gamma$-ray energy, which is emitted simultaneously with the $\alpha$-decay in the cases of odd and odd-odd nuclei \cite{asai05}. In all previous studies to determine the $\alpha$-decay energies of $^{262}$Db and $^{258}$Lr, atoms of these isotopes were collected on a plate or a film by using, for example, the gas-jet technique, and they were not implanted in the detector used. The sources of $\alpha$-decay were thus outside the detectors.  Therefore, this summing effect was not very serious in the previous studies.  Because of a shift in energy and the limited energy resolution caused by the summing effect, event ID\#11 in Group C could be assigned to the decay of $^{263}$Db $\longrightarrow $ $^{259}$Lr $\longrightarrow$ as well as that of $^{262}$Db $\longrightarrow $ $^{258}$Lr $\longrightarrow$.  The energies of $\alpha_{1}$ and $\alpha_{2}$ were measured to be 8.55 (0.09) and 8.57 (0.10) MeV, respectively. The numbers in parentheses indicate the energy resolutions in FWHM.  The time difference between the two was 2.5 s. The reported half-life and $\alpha$-decay energy of $^{263}$Db are 27 $^{+ 10}_{- 7}$ s and  8.36 MeV \cite{krats92} and those of $^{259}$Lr are 6.34 $^{+ 0.46}_{- 0.42}$ s and 8.45 MeV \cite{gregorich92}, respectively.

Three of the eight correlations in Group C (ID\#1, ID\#2 and ID\#3) are the second part ($\alpha_{2}$-$\alpha_{3}$) of the triple correlations ($\alpha_{1}$-$\alpha_{2}$-$\alpha_{3}$). All the preceding correlations ($\alpha_{1}$-$\alpha_{2}$) are in Group A, as shown in Fig.~\ref{f3}-a, along with four additional $\alpha_{1}$-$\alpha_{2}$ correlations (ID\#6--ID\#9).  Because of the genetic correlation to the $^{262}$Db $\longrightarrow $ $^{258}$Lr $\longrightarrow $ decay, these three triple correlation events in Group A are conclusively identified as the correlations of $^{266}$Bh $\longrightarrow $ $^{262}$Db $\longrightarrow $ $^{258}$Lr $\longrightarrow $ decays.  The other four correlations are identified as those of  the $^{266}$Bh $\longrightarrow $ $^{262}$Db $\longrightarrow $ decay or the $^{266}$Bh $\longrightarrow $ ($^{262}$Db $\longrightarrow $ missing) $^{258}$Lr $\longrightarrow $ decay  partly because the daughter energies, which range from 8.60 to 8.66 MeV, agree well with the adopted decay energies of $^{262}$Db and $^{258}$Lr considering the above mentioned summing effect.  The mean time differences between the $\alpha_{1}$ and $\alpha_{2}$ decays of the seven correlations in this group are calculated to be 34 $^{+ 20}_{- 10}$ s. The corresponding T$_{1/2}$ is 24 $^{+ 14}_{- 7}$ s.  This value is attributable to a mixture of the $^{266}$Bh $\longrightarrow $ $^{262}$Db $\longrightarrow $ and $^{266}$Bh $\longrightarrow $ ($^{262}$Db $\longrightarrow $) $^{258}$Lr $\longrightarrow $ decays.  Including this mixture effect, the value agrees well with the adopted half-lives of $^{262}$Db (34 $\pm$ 4 s) and $^{258}$Lr (3.92 $^{+ 0.35}_{- 0.42}$ s).  The observed $\alpha$-decay energies of $^{266}$Bh range from 9.05 to 9.23 MeV.  The correlations observed and assigned here agree well with those reported by Wilk \textit{et al.} \cite{wilk00} and Qin \textit{et al.} \cite{Qin06}.

One $\alpha_{1}$-$\alpha_{2}$ correlation of the triple $\alpha$ correlation (ID\#4) exists in Group B, as shown in Fig.~\ref{f3}-a.  The corresponding $\alpha_{2}$-$\alpha_{3}$ correlation (ID\#4) exists in Group C, suggesting that the correlations in Group B also correspond to the decay $^{266}$Bh $\longrightarrow $ $^{262}$Db $\longrightarrow $ $^{258}$Lr $\longrightarrow $.  The energies of $\alpha_{2}$ and $\alpha_{3}$ were 8.54 (0.14) and 8.69 (0.07) MeV, respectively.  These energies agree well with the reported values for $^{262}$Db and $^{258}$Lr mentioned above. The observed $\alpha$ energy of $^{266}$Bh assigned here is 8.83 MeV.  Although the value just coincides with the decay energy of $^{267}$Bh reported by Wilk \textit{et al.} \cite{wilk00}, because of the observed decay characteristics, we assigned the correlated event to the decay of $^{266}$Bh $\longrightarrow $ $^{262}$Db $\longrightarrow $ $^{258}$Lr $\longrightarrow $.

One additional correlation (ID\#14) exists in Group B, as shown in Fig.~\ref{f3}-a.  This could be also assigned to the decay of $^{266}$Bh $\longrightarrow $ $^{262}$Db $\longrightarrow $ $^{258}$Lr $\longrightarrow $ because of the similarity with the energies and decay times of the correlation assigned above.  However, based on the $\alpha_{1}$-$\alpha_{2}$ correlation analysis, assignment to the decay of $^{267}$Bh $\longrightarrow $ ($^{263}$Db $\longrightarrow $) $^{259}$Lr $\longrightarrow $ is also possible if the summing effect is considered.  Therefore, this correlation could not be assigned conclusively.

$\alpha_{1}$-$\alpha_{2}$ of one $\alpha_{1}$-$\alpha_{2}$-SF triple correlation (ID\#5) is classified in a circle marked by D in Fig.~\ref{f3}-a.  Here, the mother and daughter energies are 8.84 and 8.42 MeV, respectively.  The time difference between these decays was 12.0 s.  These values, along with the followed SF decay with a decay time of 27.2 s, are fully consistent with the decay of $^{267}$Bh $\longrightarrow $ $^{263}$Db $\longrightarrow $ $^{259}$Lr (SF) \cite{wilk00, gregorich92}. The reported T$_{1/2}$ values of $^{263}$Db and $^{259}$Lr are 27 $^{+ 10}_{- 7}$ s and 6.2 $\pm$ 0.3 s, respectively.  We then assign this correlation to the decay originating from $^{267}$Bh.

\subsection{$\alpha$-SF correlations}

\begin{figure}[t]
\begin{center}
\resizebox{0.8\textwidth}{!}{\includegraphics*{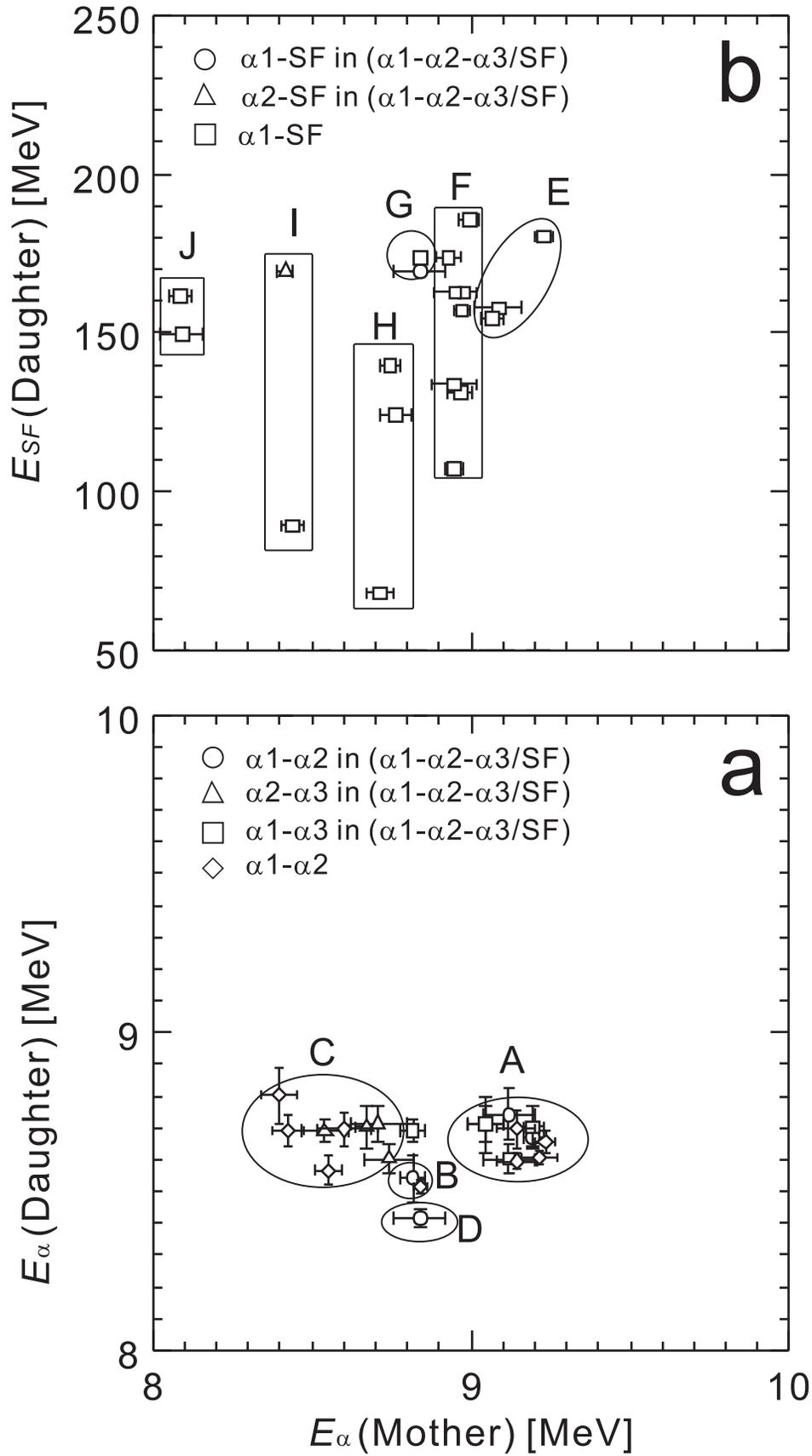}}%
\end{center}
\caption{
Two-dimensional representation of time- and position-correlated events. The lower panel (a) shows the $\alpha$-$\alpha$ correlations. The upper panel (b) shows the $\alpha$-SF correlations.  The time window was set to 300 s.
}
\label{f3}
\end{figure}

The three correlated events (ID\#15, ID\#16 and ID\#17) in Group E in Fig.~\ref{f3}-b correspond to the $\alpha_{1}$ energies of Group A (9.05--9.23 MeV).  The corresponding half-life T$_{1/2}$ deduced from the three events is 31 $^{+ 41}_{- 11}$ s, which agrees well with that of $^{262}$Db, i.e. 34 $\pm$ 4 s.  Therefore, we assign these correlated events to SF decays of $^{262}$Db fed by 9.05--9.23 MeV $\alpha$-decays of $^{266}$Bh.

Eight correlations (ID\#18--ID\#25) are observed in Group F in Fig.~\ref{f3}-b.  The $\alpha$-decay energies of the mother range from 8.93 to 8.99 MeV.  T$_{1/2}$ of the SF decay is 59 $^{+ 32}_{- 15}$ s. No $\alpha$-decays are observed in the $\alpha_{1}$-$\alpha_{2}$ correlations in this energy range, as shown in Fig.~\ref{f3}-a.  The observed T$_{1/2}$ is slightly longer than the reported  T$_{1/2}$ values of both $^{262}$Db, 34 $\pm$ 4 s, and $^{263}$Db, 27 $^{+ 10}_{- 7}$ s.  Two $\alpha$-decay events having an energy of around 8.96 MeV were reported by Qin \textit{et al.} \cite{Qin06}, and these are assigned to the decay of $^{266}$Bh $\longrightarrow $ $^{262}$Db $\longrightarrow $.  However, their experimental setup was not sensitive to the $\alpha$-SF correlations. The $\alpha$-decays having an energy of around 8.96 MeV observed in the present work fed a state in the daughter that decays  mainly by SF with T$_{1/2}$ of 59 $^{+ 32}_{- 15}$ s.  This $\alpha$-decay could be tentatively assigned to the $^{266}$Bh $\longrightarrow $ $^{262}$Db (SF) decay.  However, the possibility of assignment to the  $^{267}$Bh $\longrightarrow $ $^{263}$Db (SF) decay could not be excluded.  It should be noted that half the correlations were observed at the lowest incident energy.

One correlation (ID\#30) exists in Group G, as shown in Fig.~\ref{f3}-b, along with the $\alpha_{1}$(-$\alpha_{2}$)-SF correlation (ID\#5) that was assigned to the decay of $^{267}$Bh $\longrightarrow $ $^{263}$Db $\longrightarrow $ $^{259}$Lr $\longrightarrow $ in the present study.  The $\alpha_{1}$ energy was 8.84 MeV, and the SF decay time was measured to be 176.8 s, which agree well with those of the correlation ID\#5.  Then, it is natural to assign this correlation to the decay of $^{267}$Bh $\longrightarrow $ $^{263}$Db (SF) or $^{267}$Bh $\longrightarrow $ $^{263}$Db $\longrightarrow $ $^{259}$Lr (SF).  However, we have assigned one correlation having almost the same $\alpha_{1}$ energy to the decay of $^{266}$Bh  (Group B in Fig.~\ref{f3}-a (ID\#8)), suggesting that the correlation might still be assigned to the decay of $^{266}$Bh.

Three correlations (ID\#26, ID\#27 and ID\#28) exist in Group H in Fig.~\ref{f3}-b.  The $\alpha$-decay energy of the mother ranges from 8.71 to 8.76 MeV.  The deduced T$_{1/2}$ of the SF decay is 44 $^{+ 55}_{- 15}$ s. One decay chain having a decay energy of 8.73 MeV was assigned to the decay of $^{267}$Bh by Wilk \textit{et al.} \cite{wilk00}.  These correlations can possibly be assigned to the decay of $^{267}$Bh $\longrightarrow$ $^{263}$Db (SF) or $^{267}$Bh $\longrightarrow$ ($^{263}$Db $\longrightarrow$) $^{259}$Lr (SF).

Two correlations exist in Group I in Fig.~\ref{f3}-b.  The $\alpha$-decay energy of the mother is approximately 8.43 MeV.  One of these correlations is the $\alpha_{2}$-SF part of the triple $\alpha_{1}$-$\alpha_{2}$-SF correlation (ID\#5).  We have assigned this correlation to the decay of $^{267}$Bh $\longrightarrow $ $^{263}$Db $\longrightarrow $ $^{259}$Lr (SF).  The other correlation (ID \#30) could also be  the decay of $^{263}$Db $\longrightarrow $ $^{259}$Lr (SF).  The mean decay time of the two event is 31.6 s. The corresponding half-life is 22 $^{+ 53}_{- 9}$ s. We could tentatively assign these correlations to the decay mentioned above although the deduced SF half-life of $^{259}$Lr is slightly longer than the adopted value, 6.2 $\pm$ 0.3 s.

Two correlations (ID\#31 and ID\#32) exist in Group J in Fig.~\ref{f3}-b. The $\alpha$-decay energy of the mother is approximately 8.09 MeV.  The mean value of the decay times is 251 s.  Because the maximum correlation time of this analysis is 300 s, we could not determine the decay time of these correlations, and therefore,  we could not assign these correlations.

\subsection{9.05--9.23 MeV $\alpha$ decay of $^{266}$Bh}

An isotope of the 107th element, $^{266}$Bh, which decays by $\alpha$-emission with an energy ranging from 9.05 to 9.23 MeV, was conclusively identified by the present study.  A state in the daughter nuclei $^{262}$Db, fed by the $\alpha$-decay, decays by $\alpha$-emission and SF.  In the observed decay chains of the isotope of the 113th element, $^{278}$113, studied by the RIKEN group \cite{morita113-1, morita113-2}, one of the great-granddaughters decayed by $\alpha$-emission with a decay energy of 9.08 MeV and decay time of 2.47 s, followed by SF.  The assignment was based on an experimental result of the work done by Wilk \textit{et al.} \cite{wilk00} that reported one atom of $^{266}$Bh assigned by the sequential $\alpha$ decays $^{262}$Db $\longrightarrow $ $^{258}$Lr $\longrightarrow $.  The present work provided further confirmation of the assignment of $^{266}$Bh observed in a decay chain originating from $^{278}$113 by demonstrating the observation of the same decay energy, as well as the observation of SF decays following the relevant $\alpha$-decay of $^{266}$Bh.  

Although a decay time analysis was performed in the macro beam OFF period, because of the small counting statistics, we could only state that the half-life is longer than 1 s.

\subsection{Cross section}

We calculated an energy-averaged, inclusive cross section for the 30 correlated events assigned to the decays of $^{266}$Bh and $^{267}$Bh.  The counting efficiencies for different correlation types differed from each other.  The efficiency is unity for the $\alpha$-SF (4 events) correlations, 0.7 for the $\alpha$-$\alpha$ (12 events) and $\alpha$-$\alpha$-SF (one event) correlations and 0.5 for the $\alpha$-$\alpha$-$\alpha$ (16 events) correlations.  The counting duty was 0.725.  The total beam dose was 1.9 $\times$ 10$^{19}$, and the target thickness was 7.7 $\times$ 10$^{17}$/cm$^{2}$.  Assuming the transmission efficiency of GARIS to 0.08, we calculated the cross section to be 50 pb.

\section{Conclusion}

An isotope of the 107th element, $^{266}$Bh, that is produced by the $^{248}$Cm($^{23}$Na, 5\textit{n}) reaction was clearly identified.    The identification was based on a genetic link to the known daughter nucleus $^{262}$Db by $\alpha$-decays. The isotope $^{267}$Bh, which is a reaction product of the 4\textit{n} evaporation channel, was  also produced and identified.  A state in $^{266}$Bh, which decays by an $\alpha$-emission with the energies ranging from 9.05 to 9.23 MeV, feeds a state in $^{262}$Db, which decays by $\alpha$-emission and by SF with a previously known half-life.  The result provided a further confirmation of the production and identification of the isotope of the 113th element, $^{278}$113, studied by a research group at RIKEN, Japan.

\section*{Acknowledgements}
We are grateful to Dr. Y. Yano and Dr. M. Kase for their continuous support, encouragement and useful suggestions.  We also thank Dr. O. Kamigaito and all accelerator staff members for their excellent operation for a long period of time.  Many thanks are also due to all members of RIKEN Nishina Center for Accelerator-Based Science, for their encouragement and support.  We thank all members of RIKEN Headquarter headed by Dr. R. Noyori for their warm support. One of the authors (K. M.) thanks his late wife.  This research was partly supported by a Grant-in-Aid for Specially Promoted Research, 10992005, from the Ministry of Education, Culture, Sports, Science and Technology, Japan.

\clearpage

\renewcommand{\baselinestretch}{0.5}
\begin{sidewaystable}
\caption{Summary of decay chains observed in the reaction of $^{23}$Na on $^{248}$Cm.}
\begin{tabular}{cccrcrcrrcccccl} \hline
ID & E$_{beam}$   & Strip & E(M)    & FWHM   & E(D)     & FWHM      & dPos   & $\tau$(D) & E(GD)  & FWHM  & dPos  & $\tau$(GD) & Group & Assignment   \\
   & MeV     &      & MeV      & MeV    &     MeV  & MeV       &  mm    &   s       &  MeV   &  MeV  &  mm   &  s         &           \\ \hline \hline
 1 & 126$^a$ &   2  & 9.05     & 0.11   & 8.71$^s$ & 0.18      & $-$0.45  &  54.91    & 8.71   & 0.11  & 0.98  & 9.23       & AC    &  $^{266}$Bh $\to$ $^{262}$Db $\to$ $^{258}$Lr \\ 
 2 & 130$^b$ &  11  & 9.12$^s$ & 0.16   & 8.74$^s$ & 0.16      &  3.53  &  13.76    & 8.60   & 0.09  & $-$7.16 & 9.36       & AC    &  $^{266}$Bh $\to$ $^{262}$Db $\to$ $^{258}$Lr \\ 
 3 & 132$^a$ &   7  & 9.20     & 0.07   & 8.67     & 0.07      &  0.86  &  13.71    & 8.70$^s$ & 0.14 & $-$0.22 & 4.72      & AC    &  $^{266}$Bh $\to$ $^{262}$Db $\to$ $^{258}$Lr \\
 4 & 132$^a$ &   7  & 8.82     & 0.07   & 8.54$^s$ & 0.14      &  1.45  &  95.45    & 8.69   & 0.07  & $-$1.45 & 3.94       & BC    &  $^{266}$Bh $\to$ $^{262}$Db $\to$ $^{258}$Lr \\ 
 5 & 132$^b$ &  13  & 8.84$^s$ & 0.12   & 8.42      & 0.05     & $-$0.12  &  11.95    & 169.5$^s$ &    & $-$0.53 & 27.22      & DGI    &  $^{267}$Bh $\to$ $^{263}$Db $\to$ $^{259}$Lr \\
 6 & 130$^b$ &  3   & 9.14     & 0.12   & 8.70     & 0.12      & $-$0.06  & 66.23     &        &       &       &            & A     &  $^{266}$Bh $\to$ $^{262}$Db or $^{258}$Lr  \\ 
 7 & 132$^a$ &  6   & 9.23     & 0.07   & 8.65     & 0.07      &  0.43  & 22.04     &        &       &       &            & A     &  $^{266}$Bh $\to$ $^{262}$Db or $^{258}$Lr  \\ 
 8 & 132$^a$ &  8   & 9.14$^s$ & 0.13   & 8.60     & 0.06      &  3.50  &  7.29     &        &       &       &            & A     &  $^{266}$Bh $\to$ $^{262}$Db or $^{258}$Lr  \\
 9 & 132$^b$ & 12   & 9.22$^s$ & 0.11   & 8.61     & 0.04      & $-$0.66  & 60.40     &        &       &       &            & A     &  $^{266}$Bh $\to$ $^{262}$Db or $^{258}$Lr  \\  
10 & 130$^b$ & 10   & 8.60$^s$ & 0.17   & 8.70     & 0.10      & $-$1.72  &  6.93     &        &       &       &            & C     &  $^{262}$Db $\to$ $^{258}$Lr  \\ 
11 & 130$^b$ &  6   & 8.55     & 0.09   & 8.57     & 0.09      &  0.12  &  2.53     &        &       &       &            & C     &  $^{262}$Db $\to$ $^{258}$Lr  tentative \\ 
12 & 130$^b$ & 10   & 8.40     & 0.11   & 8.80$^s$ & 0.18      &  2.99  &  3.73     &        &       &       &            & C     &  $^{262}$Db $\to$ $^{258}$Lr  \\ 
13 & 132$^a$ &  4   & 8.43     & 0.10   & 8.69     & 0.10      & $-$0.08  &  5.69     &        &       &       &            & C     &  $^{262}$Db $\to$ $^{258}$Lr  \\ 
14 & 132$^b$ &  8   & 8.84     & 0.04   & 8.51     & 0.04      &  0.77  & 82.15     &        &       &       &            & B     &  $^{266}$Bh $\to$ $^{262}$Db  tentative \\
15 & 126$^a$ & 1    & 9.07     & 0.07   & 154.6$^s$ &          &  0.52  &  5.67  &           &       &       &            & E     &  $^{266}$Bh $\to$ $^{262}$Db  \\
16 & 130$^b$ & 9    & 9.09$^s$ & 0.15   & 157.9     &          & $-$0.56  &  5.34  &           &       &       &            & E     &  $^{266}$Bh $\to$ $^{262}$Db  \\
17 & 132$^b$ & 8    & 9.23     & 0.06   & 180.4     &          & 1.89   & 121.53 &           &       &       &            & E     &  $^{266}$Bh $\to$ $^{262}$Db            \\ 
18 & 126$^a$ & 7    & 8.99     & 0.09   & 185.8$^s$ &          &   0.16 &  8.42  &           &       &       &            & F     &  $^{266}$Bh $\to$ $^{262}$Db   tentative  \\
19 & 126$^a$ & 11   & 8.97     & 0.05   & 157.1     &          &   1.53 & 141.86 &           &       &       &            & F     &  $^{266}$Bh $\to$ $^{262}$Db   tentative  \\ 
20 & 126$^a$ & 12   & 8.95$^s$ & 0.13   & 162.8     &          &  $-$1.56 & 68.35  &           &       &       &            & F     &  $^{266}$Bh $\to$ $^{262}$Db   tentative  \\  
21 & 126$^a$ & 7    & 8.93     & 0.08   & 173.9$^s$ &          &   0.61 & 84.30  &           &       &       &            & F     &  $^{266}$Bh $\to$ $^{262}$Db   tentative  \\ 
22 & 130$^b$ & 7    & 8.97     & 0.08   & 131.1     &          &  $-$1.20 & 43.99  &           &       &       &            & F     &  $^{266}$Bh $\to$ $^{262}$Db   tentative  \\  
23 & 132$^a$ & 1    & 8.95     & 0.06   & 107.5     &          &  $-$0.06 & 151.36 &           &       &       &            & F     &  $^{266}$Bh $\to$ $^{262}$Db   tentative  \\ 
24 & 132$^b$ & 13   & 8.98     & 0.04   & 162.8     &          & $-$0.72  & 156.99 &           &       &       &            & F     &  $^{266}$Bh $\to$ $^{262}$Db   tentative  \\
25 & 132$^b$ & 10   & 8.95$^s$ & 0.14   & 133.8     &          &   3.05 & 26.85  &           &       &       &            & F     &  $^{266}$Bh $\to$ $^{262}$Db   tentative  \\ 
26 & 126$^a$ & 4    & 8.76     & 0.10   & 124.3$^s$ &          &   0.14 & 112.21 &           &       &       &            & H     &  $^{267}$Bh $\to$ $^{263}$Db   tentative  \\ 
27 & 130$^b$ & 10   & 8.71     & 0.08   & 68.2      &          &   0.26 & 5.38   &           &       &       &            & H     &  $^{267}$Bh $\to$ $^{263}$Db  tentative\\ 
28 & 132$^b$ & 11   & 8.75     & 0.07   & 139.9$^s$ &          &  $-$0.49 & 55.57  &           &       &       &            & H     &  $^{267}$Bh $\to$ $^{263}$Db  tentative \\ 
29 & 132$^b$ & 10   & 8.44     & 0.07   & 89.4      &          &  0.64  & 35.96  &           &       &       &            & I     &  $^{263}$Db or $^{258}$Lr  \\
30 & 130$^b$ & 12   & 8.84     & 0.04   & 173.8$^s$ &          &   0.76 & 176.77 &           &       &       &            & G     &  $^{267}$Bh $\to$ $^{263}$Db or $^{259}$Lr  \\
31 & 132$^a$ & 7    & 8.09     & 0.07   & 161.7$^s$ &          &  $-$1.52 & 294.39 &           &       &       &            & J     &   not assigned \\
32 & 132$^b$ & 14   & 8.09$^s$ & 0.13   & 164.8$^s$ &          &   0.28 & 208.30 &           &       &       &            & J     &   not assigned \\
\hline
\end{tabular}
$^{a}$ B$\rho$ of GARIS was set to 2.19 \\
$^{b}$ B$\rho$ of GARIS was set to 2.07 \\
$^{s}$ Sum of PSD and SSD signals \\
\end{sidewaystable}
\renewcommand{\baselinestretch}{1.0}

\clearpage

\clearpage

\clearpage

\end{document}